\documentclass[plb,reprint]{revtex4-1}
\usepackage{amssymb,amsfonts,amsmath,graphicx}

\usepackage{amssymb,amsmath,epsfig,todonotes}

\newcommand{\eeq}{\end{equation}}
\newcommand{\beq}{\begin{equation}}

\begin{document}

\title{Counting String Theory Standard Models}



\author{Andrei Constantin}
\affiliation{Department of Physics and Astronomy, Uppsala University, SE-751 20, Uppsala, Sweden}
\email[]{andrei.constantin@physics.uu.se}

\author{Yang-Hui He}
\affiliation{Department of Mathematics, City, University of London, EC1V0HB, UK,}
\affiliation{Merton College, University of Oxford, OX14JD, UK,}
\affiliation{School of Physics, NanKai University, Tianjin, 300071, China}
\email[]{hey@maths.ox.ac.uk}

\author{Andre Lukas}
\affiliation{Rudolf Peierls Centre for Theoretical Physics, Oxford University, 1 Keble Road, Oxford, OX1 3NP, U.K.}
\email[]{lukas@physics.ox.ac.uk}

\date{\today}

\begin{abstract}\vskip 3mm\noindent
We derive an approximate analytic relation between the number of consistent heterotic Calabi-Yau compactifications of string theory with the exact charged matter content of the standard model of particle physics and the topological data of the internal manifold: the former scaling exponentially with the number of K\"ahler parameters. 
This is done by an estimate of the number of solutions to a set of Diophantine equations representing constraints satisfied by any consistent heterotic string vacuum with three chiral massless families, and has been computationally checked to hold for complete intersection Calabi-Yau threefolds (CICYs) with up to seven K\"ahler parameters. 
When extrapolated to the entire CICY list, the relation gives $\sim\!\!10^{23}$ string theory standard models; for the class of Calabi-Yau hypersurfaces in toric varieties, it gives $\sim\!\!10^{723}$ standard models. 
\end{abstract}


\maketitle


\section{Introduction and Summary}
It is generally believed that string compactifications that have the exact charged matter content of the standard model of particle physics (and no other charged matter except moduli) are few in number. The purpose of this letter is to show that, although such compactifications may be rare and hard to find, their number is substantial. Admittedly, this bias has come in the past from the difficulty to construct phenomenologically viable compactifications. However, since the birth of string phenomenology in Ref.~\cite{Candelas:1985en}, from the advent of the first standard-like string model \cite{Greene:1986ar}, to the first exact particle spectrum directly derived from a string compactification \cite{Braun:2005ux,Braun:2005nv,Braun:2006ae,Bouchard:2005ag}, to the first result \cite{Anderson:2009mh} from algorithmic heterotic compactification \cite{Anderson:2007nc}, until the comprehensive computer scan of Refs.~\cite{Anderson:2011ns,Anderson:2012yf,Anderson:2013xka,Constantin:2015bea}, as well as the various statistical perspectives on the heterotic landscape \cite{stats} (cf.~\cite{Gmeiner:2005vz} in Type II and beyond \cite{Douglas:2003um}), there has been much progress.

While it can be specified at different levels of sophistication, for this letter a ``string standard model" is a model with a massless spectrum which is exactly that of the minimally supersymmetric standard model (MSSM), plus any number of massless modes (moduli fields) {\em uncharged} under the standard model gauge group. 

The general strategy of heterotic string phenomenology is to consider a smooth, compact Calabi-Yau threefold (CY), say $X$, with a non-trivial fundamental group~$\Gamma$, together with a holomorphic, (poly-)stable vector bundle $V$ over $X$, typically with structure group $SU(5)$ or $SU(4)$.
Subsequently, a $\Gamma$-Wilson line can break the Grand Unified Theory (GUT) group, typically $SU(5)$ or $SO(10)$, to the MSSM group and the $\Gamma$-equivariant cohomology of $V$ as well as its appropriate tensor powers correspond to the MSSM particles. 

From an algorithmic point of view, one can (1) take manifolds $\tilde{X}$ from existing databases, most of which are simply connected, then search for discrete, freely acting symmetry groups $\Gamma$ on~$\tilde{X}$, and consider the quotient $X \simeq \tilde{X} / \Gamma$ with fundamental group $\Gamma$; 
(2) construct and classify families of $\Gamma$-equivariant bundles $V$ on $X$, ensure stability,  and then compute the relevant cohomology groups;
and (3) scan through the results to look for exact MSSM particle content. Much of these can be implemented on a computer.

The most extensively used databases of manifolds are the Complete Intersection Calabi-Yau three-folds (CICYs) embedded in products of projective spaces of around 8000 manifolds \cite{cicy,huebsch} as well as the Kreuzer-Skarke (KS) dataset of Calabi-Yau hypersurfaces embedded in four-dimensional toric varieties of around half-billion \cite{ks,palp,Altman:2014bfa}.
The comprehensive scan of \cite{Anderson:2013xka} was performed on CICY manifolds with a number of K\"ahler parameters less than $h^{1,1}(X) = 6$ and vector bundles constructed from sum of line bundles. A total of about $35,000$ $SU(5)$ heterotic line bundle models were found, all with the right field content to induce low-energy standard-like models. 

It is obviously important to have a count of MSSM models expected within string theory. 
While ours is still a relatively unrefined notion of the standard model, even counting at this level is not easy since it requires information on all vector bundles and their cohomology on CYs which is not available in any systematic form. An exception to this rule is the class of vector bundles that split into a sum of line bundles. Holomorphic line bundles are classified by their first Chern class, which can be expressed in terms of $h^{1,1}(X)$ integers. As such, line bundles can be enumerated. 

Moreover, there is enough evidence that indicates the existence of analytic formulae for the ranks of line bundle valued cohomology groups in terms of the line bundle integers \cite{Constantin:2018hvl, Klaewer:2018sfl, Buchbinder:2013dna, Constantin:2013}. Finally, line bundle sums offer an accessible window into the moduli space of non-abelian bundles \cite{Buchbinder:2013dna, Buchbinder:2014qda, Buchbinder:2014sya}: if a line bundle sum corresponds to a standard-like model, then usually it can be deformed into non-abelian bundles that also lead to standard-like models. 

Our model building experience for such (rank five) line bundle models on CICYs suggests that a significant number of consistent models with the correct chiral asymmetry will descend to standard models after dividing by the freely acting discrete symmetry group~$\Gamma$. 
Moreover, by far the most frequent symmetry is $\mathbb{Z}_2$. This suggests that an indication of the number of standard models should be provided by counting the consistent upstairs line bundle models with chiral asymmetry $6$, relevant for $\mathbb{Z}_2$ symmetries.

We start our analysis by outlining the constraints on the compactification data that guarantee an exact MSSM spectrum. For a fixed manifold, most of these constraints take the form of Diophantine equations and inequalities, where the unknown variables are the line bundle integers. For the class of CICY manifolds with less than six K\"ahler parameters this system was solved in Ref.~\cite{Anderson:2013xka} by explicitly checking every possible line bundle sum. We augment this dataset of line bundle models with results for $7$ new manifolds with $h^{1,1}(X) = 6,7$. 

These scans suggest a simple rule:  the number of line bundle models increases roughly by an order of magnitude with every increment of $h^{1,1}(X)$ by one. However, it is difficult to test this relation for larger values of $h^{1,1}(X)$ due to computer limitations. Instead, we estimate the number of solutions to the Diophantine system of constraints using a result from the mathematical literature \cite{HeathBrown}. For this to hold, we define a bound on the line bundle integers in terms of topological data of the manifold. 

Finally, we come back to the empirical dataset of line bundles and correlate the number of solutions not only with $h^{1,1}(X)$, but also with a number of topological invariants, constructed from the intersection form and the second Chern class of the CY manifold, that display little variation with increasing $h^{1,1}(X)$. Extrapolating the multi-linear regression to the maximal value of $h^{1,1}(X)$ found in the CICY dataset, we estimate a total of $N_{\rm CICY}\simeq 10^{23}$ line bundle models (we owe the expression  ``a mole of models'' to 
Tristan H\"ubsch)
while for the manifolds in the Kreuzer-Skarke list we expect $N_{\rm KS}\simeq 10^{723}$ line bundle models. 

~\\

\paragraph*{{\bf Acknowledgments}}
We are grateful to K.~Dienes and T.~H\"ubsch for valuable comments on the draft.
AC and AL would like to thank the Mainz Institute for Theoretical
Physics for hospitality during part of the completion of this project.
AL is partially supported by the EPSRC network grant EP/N007158/1.
YHH thanks STFC for grant ST/J00037X/1.

~\\



\section{Counting Line Bundle MSSMs}
The models of interest for our count have an exact MSSM particle content and are constructed from heterotic compactifications on a smooth, compact Calabi-Yau threefolds $X$ endowed with slope-zero, poly-stable direct sums of line bundles. Let $h$ denote the Picard number of $X$, $h:=h^{1,1}(X)$, and choose an integral basis of $H^2(X)$ denoted by $\{J_i\}$, where $i=1,\ldots ,h$.
In this basis, let the second Chern class of $X$ be $c_{2,i}$ and the triple intersection numbers be $d_{ijk}=\int_XJ_i\wedge J_j\wedge J_k$. 
Line bundles $L\rightarrow X$ with first Chern class $c_1(L)=k^i J_i$ are denoted by $L={\cal O}_X({\bf k})$.
The counting problem can then be formulated as follows:\\[2mm]
{\bf PROBLEM: } 
What is the number $N = N(h,c_{2,i},d_{ijk})$ of rank five line bundle sums $V=\oplus_{a=1}^5 L_a$, where $L_a={\cal O}_X({\bf k}_a)$ satisfying the following constraints:
\begin{description}
\item[$E_8$ embedding]  $c_1(V) =\sum\limits_{a=1}^5 k^i_a\stackrel{!}{=}0$ for all $i=1,\ldots ,h;$
\item[Anomaly cancellation] 
$$c_{2,i}(V)=-\frac{1}{2}d_{ijk}\sum_a k_a^jk_a^k\stackrel{!}{\leq}c_{2,i}\text{ for all }i=\ldots ,h;$$
\item[Supersymmetry/Zero Slope] there is a common solution $t^i$ to the vanishing slopes 
$$\mu(L_a)=d_{ijk}k_a^it_a^jt_a^k\stackrel{!}{=}0\text{ for }a=1,\ldots ,5$$ such that $J=t^iJ_i\in$ interior of the Kahler cone;
\item[Particle generations] the chiral asymmetry is six, i.e. 
$${\rm ind}(V)=\frac{1}{6}d_{ijk}\sum_ak_a^ik_a^jk_a^k\stackrel{!}{=}-6.$$
\end{description}
We emphasize that $N$ is a function of the prescribed Hodge number $h$, the second Chern class $c_{2,i}$, as well as the triple intersection numbers $d_{ijk}$ of $X$.

A few remarks should be added and the reader is also referred to Sec.~4 of  Ref.~\cite{Anderson:2013xka} for further details on the above constraints. 
The above counting is only concerned with $SU(5)$ bundles $V$ for the following reason.
From a group theoretic point of view \cite{Slansky:1981yr}, there are many ways to break the GUT group to the MSSM group using an appropriate discrete Wilson line, for example, the exact MSSM spectrum of \cite{Braun:2005nv} was achieved with a $\mathbb{Z}_3 \times \mathbb{Z}_3$ Wilson line from an $SO(10)$ GUT group.
However, CY manifolds $\tilde{X}$ with a large freely acting discrete symmetry group $\Gamma$  are quite rare.
This can be seen, for instance, from the complete classification of freely-acting \cite{Braun:2010vc} and residual \cite{Candelas:2017ive,Lukas:2017vqp} symmetries on all CICYs, or from the KS dataset of hypersurfaces in toric Fano fourfolds \cite{He:2013ofa,Braun:2017juz}.

Therefore, {\it generically}, it is expected that Calabi-Yau manifolds with a small fundamental group $\pi_1(X)$, should far exceed in number those with a large $\pi_1(X)$ (this should be contrasted with the relative paucity of Calabi-Yau manifolds of small Hodge numbers \cite{Candelas:2008wb,Candelas:2016fdy, Constantin:2016xlj}). The smallest possible $\pi_1(X)$ that breaks the $SU(5)$ GUT group to the Standard Model gauge group is $\Gamma = \mathbb{Z}_2$ and this setup is expected to dominate.
Now, in $SU(5)$ (commutant of the $SU(5)$ of the bundle in $E_8$) GUTs, the {$\overline{\bf 10}$} representation corresponds entirely to anti-families which we desire to be absent. Under the branching of $E_8$ to $SU(5)$, this corresponds to the condition that $h^2(X,V)=0$, so that stability (implying that $h^0(X,V) = h^3(X,V) = 0$) in conjunction with the index theorem gives 
\begin{equation}
-3 |\Gamma| = {\rm ind}(V) = \sum\limits_{i=0}^3 h^i(X,V) = -h^1(X,V) \ ,
\end{equation}
Thus, we impose ${\rm ind}(V) = -h^1(X,V) = -6$.
 
\subsection{Preliminary Count}\label{s:rough}
\begin{figure}[h!!!]
\begin{center}
\includegraphics[width=8cm]{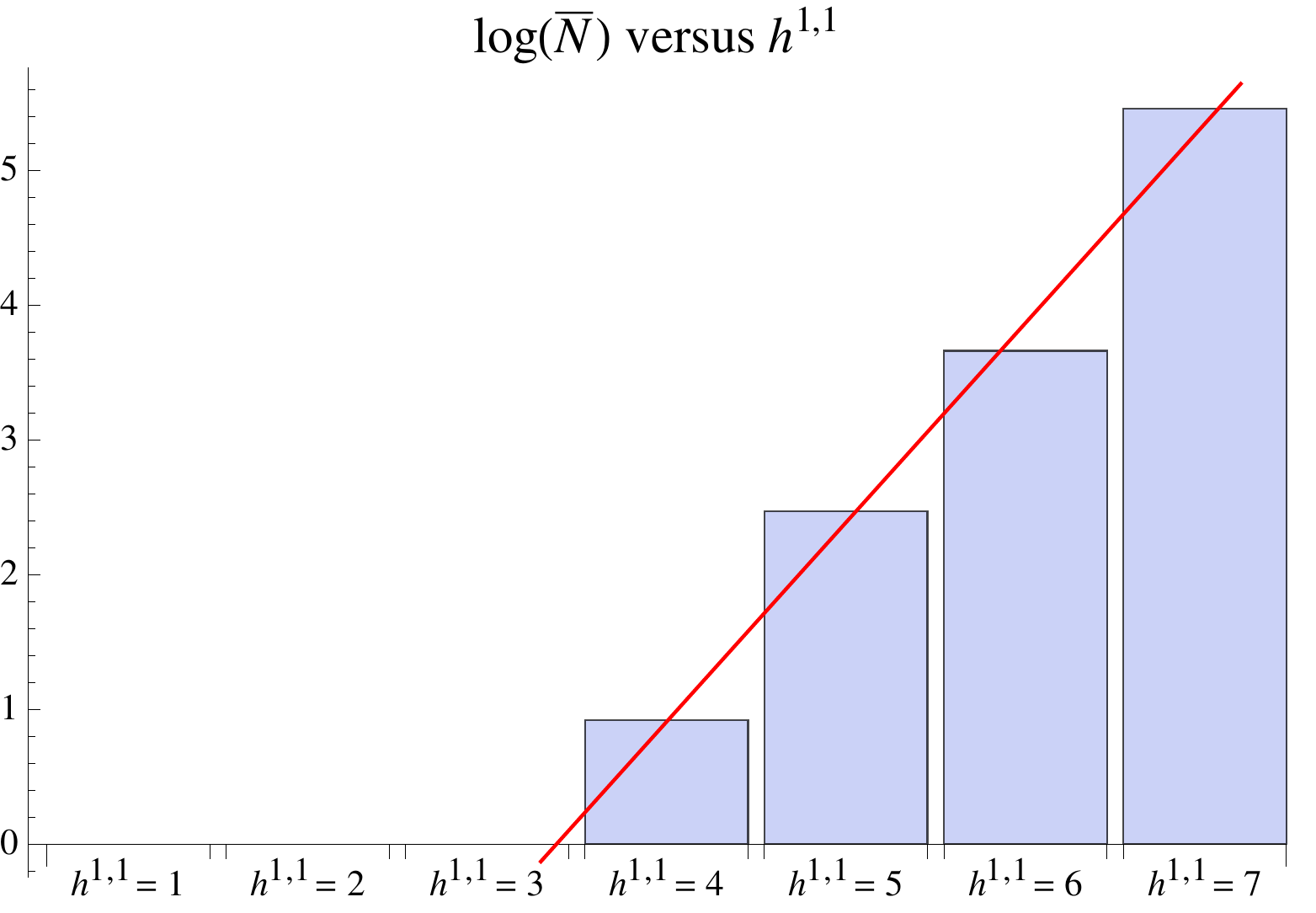}
\caption{\sf The logarithm of $\bar{N}$, the average number of models per CY, as a function of $h=h^{1,1}(X)$, taken from the data in Table~\ref{tab1}. The red line is a linear fit to the data.}\label{fig1}
\vspace{-21pt}
\end{center}
\end{figure}

For the subset of favourable CICYs with $h\leq 6$, the number $N$ was determined by the computer scan \cite{Anderson:2013xka} and we have extended this scan to include manifolds for $h^{1,1}(X)=7$ as well  as some non-favourable manifold for $h^{1,1}\leq 7$ which were previously discarded. The cumulative and average number of models found for each Picard number $h$ is summarised in Table~\eqref{tab1}.
Note there are no viable models for $h=1,2,3$ since the supersymmetry (slope zero) conditions are too constraining for those cases.
\begin{table}[h!!!]
\begin{center}
\begin{tabular}{|c||c|c|c|c|}\hline
$h=h^{1,1}(X)$&4&5&6&7\\[4pt]\hline\hline
number of CYs&9&10&8&5\\[4pt]\hline
number of models&75&2949&36692&1428856\\[4pt]\hline
~number, $\bar{N}$, of models per CY~~&8.3&294.9&4586.5&285771.0\\[4pt]\hline
\end{tabular}
\caption{\sf Number of models found in a computer scan for favourable CICYs with $4\leq h\leq 7$ for each $h$.}\label{tab1}
\vspace{-21pt}
\end{center}
\end{table}
\vskip 2mm\noindent
A simple approach is to assume that the main dependence of $N$ is on $h$, and to neglect the possible effects of $c_{2,i}$ and $d_{ijk}$.
The average number, $\bar{N}=\bar{N}(h)$ of models per CY as a function of $h$, taken from the last row of Table~\ref{tab1}, has been plotted (logarithmically) in Fig.~\ref{fig1}.

A linear fit to this data (which corresponds to the red line in Fig.~\ref{fig1}) leads to
\begin{equation}\label{crudeFit}
 \log(\bar{N}(h))\simeq -5.0+1.5\, h\; .
\end{equation} 
The largest known Picard number of any CY threefold is $h_{\rm max}=491$, which appears within the KS data set, and the largest value within the CICY list is $h_{\rm CICY}=19$. 
Using \eqref{crudeFit} to boldly extrapolate to those values we find
\begin{equation}
\bar{N}(h_{\rm CICY})\simeq 10^{23}\;,\qquad  \bar{N}(h_{\rm max})\simeq 10^{721}\; . \label{extra1}
\end{equation} 
Clearly, these numbers are quite dramatic, even if we restrict ourselves to the CICYs. The predicted number of standard models even within this set is significantly larger than can be currently stored, let alone found by a scan. 
However, the method so far is quite crude and the extrapolation to large $h$ adventurous. 
\begin{figure}[h!!!]
\begin{center}
\includegraphics[width=8cm]{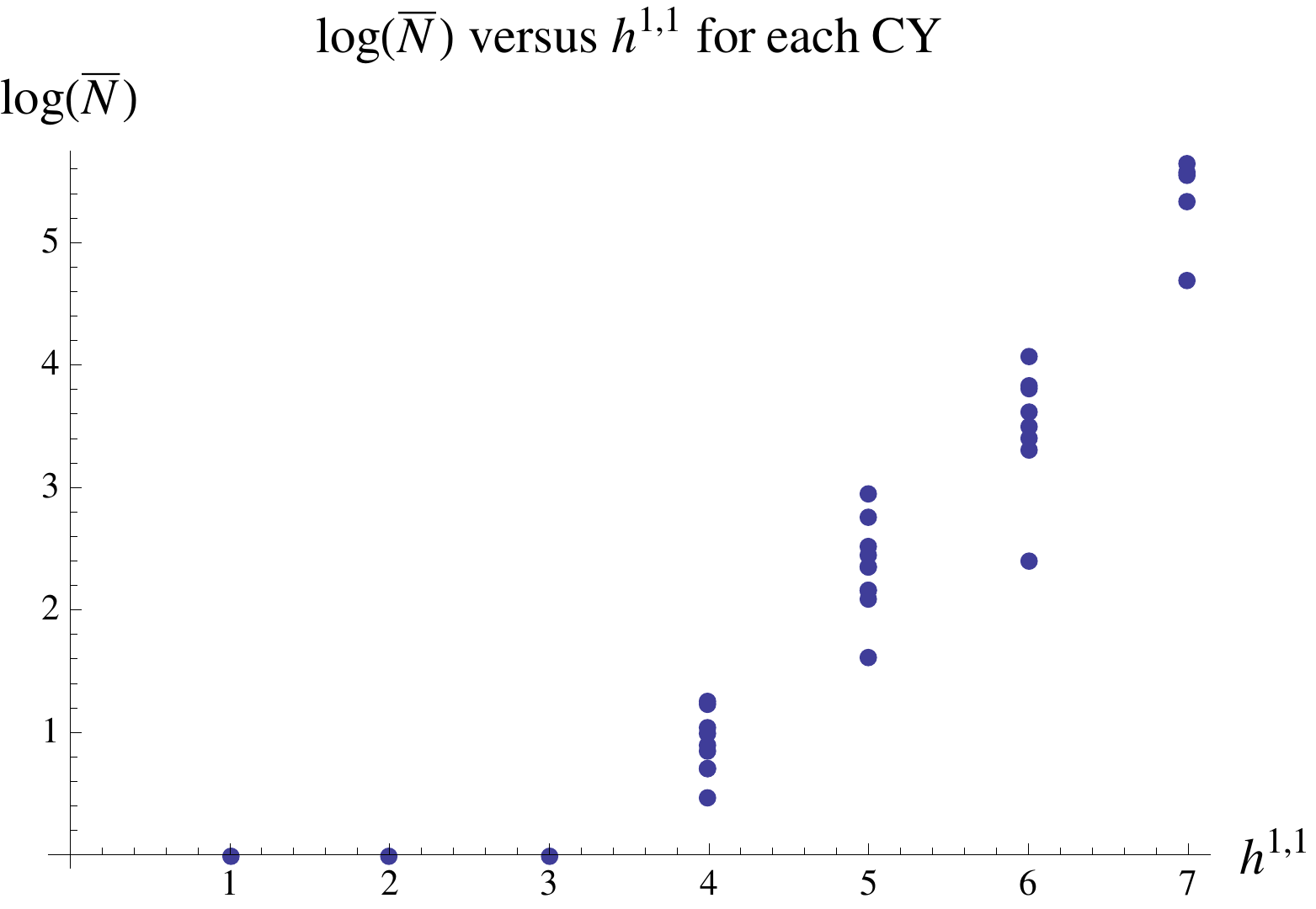}
\caption{\sf The logarithm of $N$, the number of models for each of the CICY manifolds, as a function of $h=h^{1,1}(X)$.}
\label{fig2}
\end{center}\vskip -5mm
\end{figure}
To see some of the problems, consider Fig.~\ref{fig2} which shows the number of models as a function of $h$ for each CICY, rather than the average over all CICYs with the same $h$, as in Fig.~\ref{fig1}.
The variation within given $h$ can be seen to be considerable - clear indication that there is a strong dependence of $N$ on $c_{2,i}$ and $d_{ijk}$, in addition to $h$.

\subsection{Some Theoretical Considerations}\label{s:th}
While the computer scan gives a finite number of models in each case, it is actually not easy to prove that $N$ is finite. 
A succinct argument was presented in Ref.~\cite{Constantin:2015bea} based on the moduli space metric 
\begin{equation}\label{Gij}
 G_{ij}=-3\left(\frac{\kappa_{ij}}{\kappa}-\frac{3}{2}\frac{\kappa_i\kappa_j}{\kappa^2}\right)\; ,
\end{equation}
where $\kappa=d_{ijk}t^it^jt^k$, $\kappa_i=d_{ijk}t^jt^k$ and $\kappa_{ij}=d_{ijk}t^k$, with $t^i$ being the K\"ahler moduli.
We note that the slope zero conditions can be expressed as $\kappa_ik_a^i=0$ so that
\begin{align}
\nonumber
 0 & \leq\sum_a{\bf k}_a^TG\,{\bf k}_a=-\frac{3}{\kappa}d_{ijk}\sum_ak_a^ik_a^jt^k=\frac{6}{\kappa}t^ic_{2,i}(V) \\
   & \leq \frac{6}{\kappa}t^ic_{2,i}(TX)\leq \frac{6}{\kappa}|{\bf t}|\,|c_{2,i}(TX)|\; .
\end{align} 
Then, introducing the scale-invariant modified metric
$
 \tilde{G}=\frac{\kappa}{6|{\bf t}|}G
$,
we get the bound 
\begin{equation}
\sum_a{\bf k}_a^T\tilde{G}\,{\bf k}_a\leq |c_{2,i}(TX)|\; . \label{bound}
\end{equation}
This by itself does unfortunately not bound the vectors ${\bf k}_a$ since the metric $\tilde{G}$ might become singular at the boundary of the K\"ahler cone. However, if we require that we stay in a ``physical" region of the K\"ahler cone where all $t^i>1$ (so that the supergravity approximation is valid) and the volume $\kappa$ is bounded from above (so that we are not de-compactifying) then the eigenvalues of $\tilde{G}$ are bounded from below by a strictly positive number. 

Eq.~\eqref{bound} then implies that the length $\sum_a|{\bf k}_a|^2$ is bounded from above and, hence, that there is only a finite number of possible integer vectors ${\bf k}_a$. More quantitative statements depend very much on the specific example but what can be said is that the length of the ${\bf k}$ vectors is bounded by a radius $R$ which roughly scales as
\begin{equation}
 R\sim\left(\frac{c_2}{d}\right)^{1/2}\; , \label{Rdef}
\end{equation}
where $c_2$ and $d$ are typical values of $c_{2,i}$ and $d_{ijk}$. 

Apart from the slope zero conditions the constraints on the line bundles can be written as a system of Diophantine equations
\begin{align}
\nonumber  \sum_{a=1}^5k^i_a&\stackrel{!}{\,=\,}0\;,\qquad 
\nonumber -\frac{1}{2}d_{ijk}\sum_a k_a^jk_a^k\stackrel{!}{\,=\,}c_{2,i}\;,\\
&\frac{1}{6}d_{ijk}\sum_ak_a^ik_a^jk_a^k\stackrel{!}{\,=}-6\;,
\end{align}
for $i=1,\ldots ,h$. Here we have replaced the inequality in the anomaly condition with an equality, assuming that  the bulk of the contribution comes from line bundles with the largest allowed integers. We can homogenise these equations (by introducing one additional coordinate) and think of them as a set of equations in $\mathbb{P}^n$, where $n=5\,h$. Assuming they provide a complete intersection, $Z$, its dimension is given by $m=n-2\,h-1=3\,h-1$.

\begin{figure}[h]
\begin{center}
\includegraphics[width=8cm]{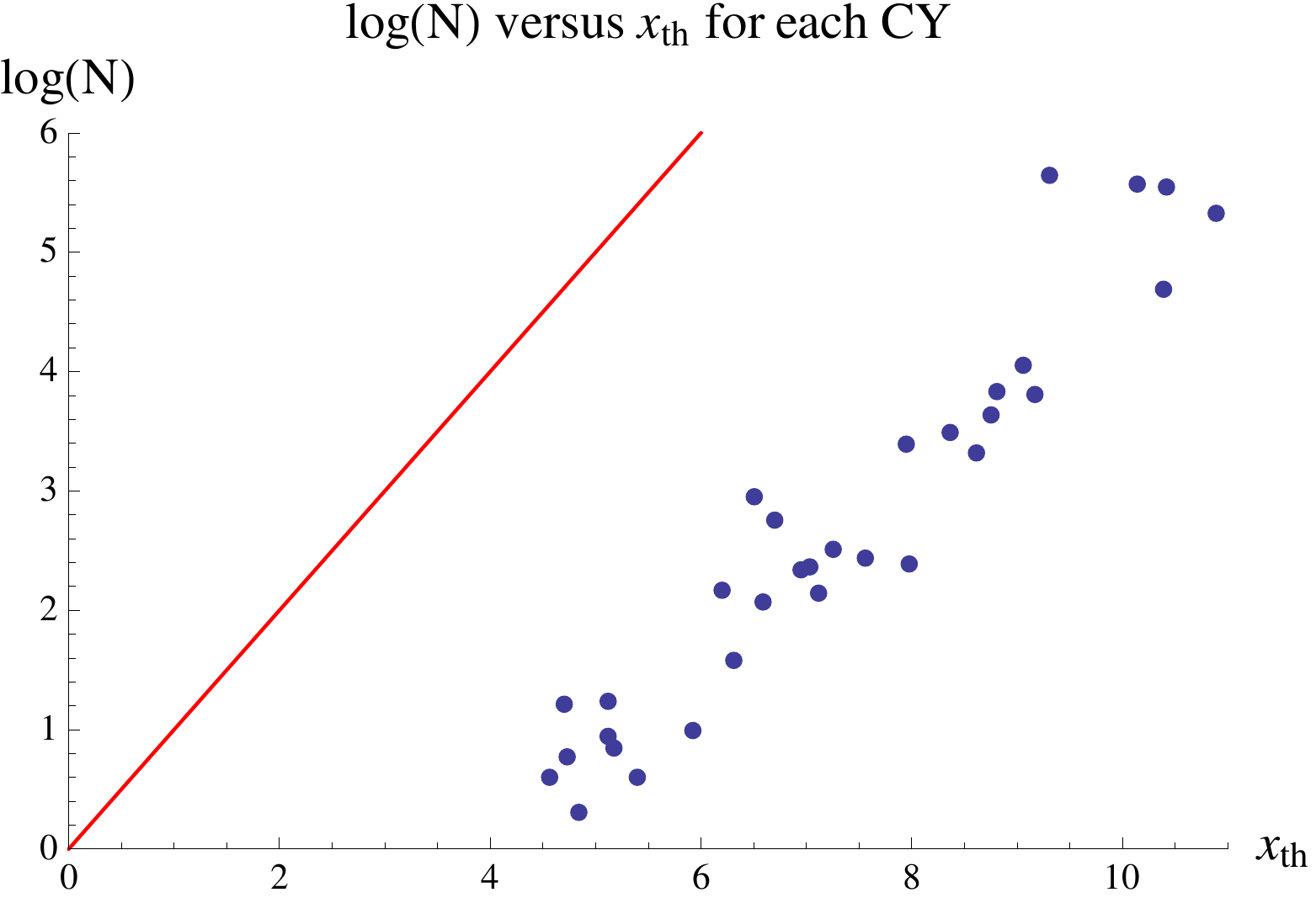}
\caption{\sf The logarithm of $N$ from the computer scan of CICYs versus the "theoretical" upper bound $x_{\rm th}$, as defined in Eq.~\eqref{xth}.}\label{fig3}
\end{center}\vskip -5mm
\end{figure}

 Ref.~\cite{HeathBrown} provides an upper bound for the number of rational points $N_Z(B)$ within a box of size $B$ (where the size is measured by the maximum norm) on $Z$ which is given by $N_Z(B)\ll B^m$. Using the above radius $R$ as an upper bound for $B$ we find
\begin{equation}
 N\ll R^{3h-1}\sim\left(\frac{c_2}{d}\right)^{(3h-1)/2}=:x_{\rm th}\; . \label{xth}
\end{equation} 
The comparison between this upper bound and the results from the computer scan on CICY manifolds is shown in Fig.~\ref{fig3}.
With all the data points well below the red line it is clear that Eq.~\eqref{xth} indeed provides an upper bound but it is equally obvious that this upper bound is rather weak. There are a number of possible reasons for this. First, the result of Ref.~\cite{HeathBrown} is only an upper bound (which, in addition, counts rational rather than integer points). Second, the radius $R$ from Eq.~\eqref{Rdef} is a crude estimate and is, by itself, only an upper bound on the size $B$ of the box considered in Ref.~\cite{HeathBrown}.


\subsection{A more sophisticated count}
Our count in \S\ref{s:rough}, based on considering only the dependence of $N$ on the Picard number $h$ is clearly somewhat unrefined, while the theoretical upper bound $x_{\rm th}$ from \S\ref{s:th} is clearly too weak to allow for a meaningful extrapolation to larger values of $h$. 
In this subsection, we seek a more sophisticated equation for $N$ as a function of $h$, $c_{2,i}$ and $d_{ijk}$, drawing inspiration from the above discussions.

\begin{figure}[h!!!t!]
\vskip 7mm
\begin{center}
\includegraphics[width=8cm]{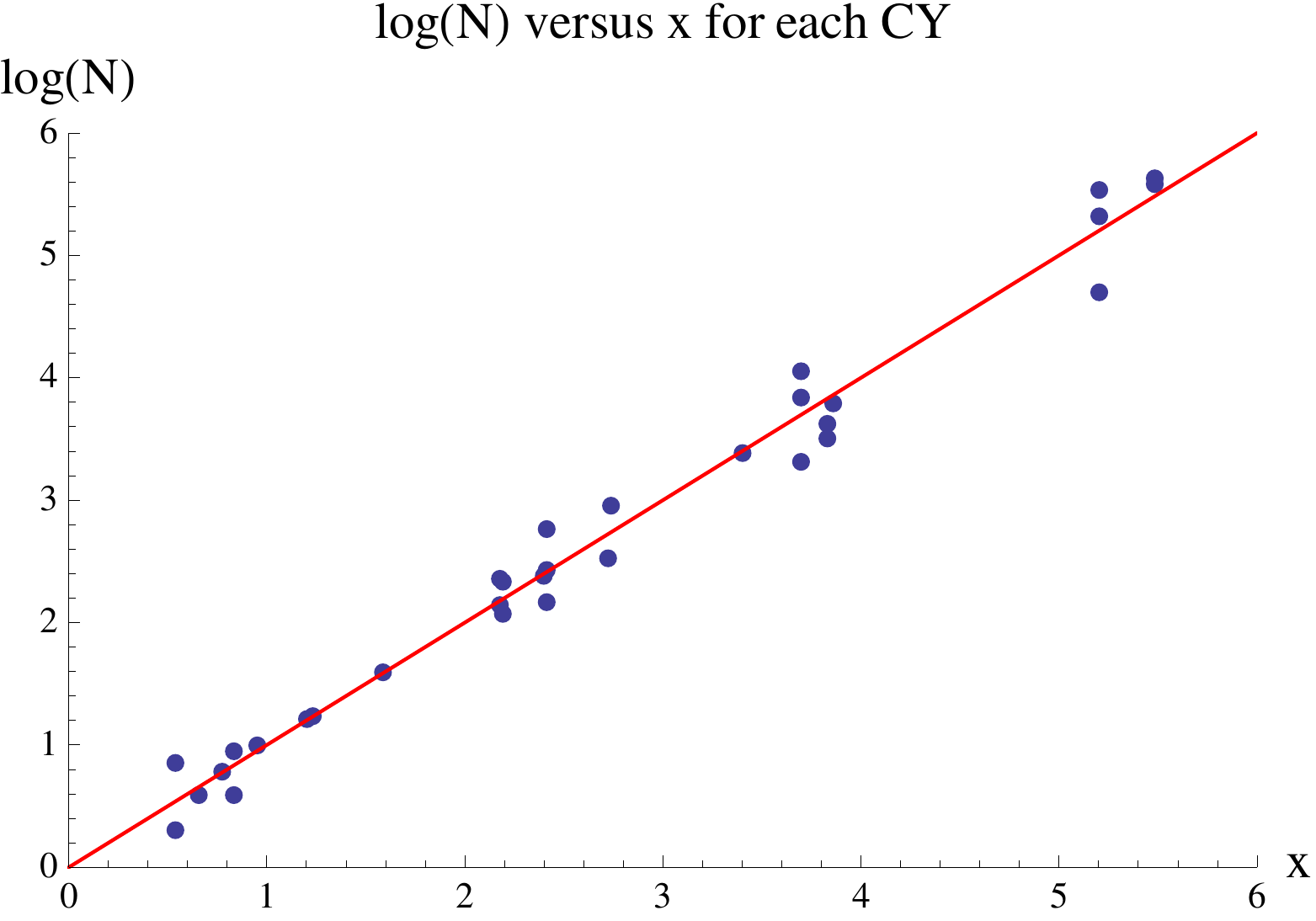}
\caption{\sf The logarithm of $N$ from the computer scan of CICYs versus the quantity $x$ in Eq.~\eqref{x}, where the constants $A_i$ and $B_i$ have been determined to provide the best fit.}\label{fig4}
\end{center}
\vskip -8mm
\end{figure}

There is an obvious difficulty of writing down even an ansatz for $N$ as a function of $c_{2,i}$ and $d_{ijk}$:  both of these quantities are basis-dependent (on a choice of integral basis $J_i$ for $H^2(X)$), but $N$ clearly cannot depend on such a  choice of basis. This means we should think about basis-independent quantities which can be constructed from $c_{2,i}$ and $d_{ijk}$. 

Unfortunately, both quantities have ``all indices down" so there is no invariant which can be obtained by a simple contraction of indices, given that the only available metric, $G$ from Eq.~\eqref{Gij}, is moduli-dependent.
This problem has been encountered before, in the context of practical applications of  Wall's theorem, and a solution has been proposed in Ref.~\cite{huebsch}, p.~174. 

From the intersection form $\lambda(x,y,z):=\int_Xx\wedge y\wedge z \in \mathbb{Z}_{\geq 0}$, completely symmetric in $x,y,z$,  the following invariants can be constructed:
\begin{align}
\nonumber
 \ell_1&={\rm gcd}\{\lambda(x,y,z)\,|\, x,y,z\in H^2(X,\mathbb{Z})\}\\
\nonumber
 \ell_2&={\rm gcd}\{\lambda(x,y,y)\,|\, x,y\in H^2(X,\mathbb{Z})\}\\
 \ell_3&={\rm gcd}\{\lambda(x,x,x)\,|\, x\in H^2(X,\mathbb{Z})\} \ .
 \end{align}
Furthermore, combining the intersection form and $c_2=c_2(TX)$ we can define the form 
$\Lambda(x,y,z,w)=(\lambda(x,y,z)c_2(w)+\mbox{3 permutations})$ which gives rise to the invariants
\begin{align}
\nonumber
 \ell_4&={\rm gcd}\{\Lambda(x,y,z,t)\,|\, x,y,z,t\in H^2(X,\mathbb{Z})\}\\
 \nonumber
 \ell_5&={\rm gcd}\{\Lambda(x,y,z,z)\,|\, x,y,z\in H^2(X,\mathbb{Z})\}\\
 \nonumber
 \ell_6&={\rm gcd}\{\Lambda(x,y,y,y)\,|\, x,y\in H^2(X,\mathbb{Z})\}\\
 \ell_7&={\rm gcd}\{\Lambda(x,x,x,x)\,|\, x\in H^2(X,\mathbb{Z})\} \ .
\end{align} 
Ref.~\cite{huebsch} also provides a practical way of computing these invariants which involves a scan over only a finite subset of $H^2(X,\mathbb{Z})$, so that they can be worked out from $d_{ijk}$ and $c_{2,i}$. 

Combining the approaches of \S\ref{s:rough} and \S\ref{s:th} and using invariants $\ell_{i=1,\ldots,7}$ a plausible ansatz for $x := \log N$ is
\begin{equation}
  x=\sum_{i=1}^7(A_i+B_i\,h)\log  \ell_i\; , \label{x}
\end{equation}
where $A_i$ and $B_i$ are constants to be determined by regression.  
Using the data from the CICY scan, together with the Hodge numbers $h$ and the invariants $\ell_i$ for each CICY involved, the best fit values of $A_i$ and $B_i$ are
\begin{equation}
\begin{array}{lll}
(A_i)&\simeq& (12.2, 0, 11.6, -3.5, -11.7, 1.6, 1.8)\\
(B_i) &\simeq& (-2.3, 0, -1.9, 0.9, 2.0, -0.7, 0.2)\; .
\end{array}
\label{AB}
\end{equation}

The comparison of this fit with the data is provided in Fig.~\ref{fig4}. Each point corresponds to a CICY with $x$ is the value computed from the RHS of Eq.~\eqref{x}, using the values in \eqref{AB} for $A_i$, $B_i$ and $\log(N)$ is the value of standard models on this CICY found by the computer scan. The red line is the diagonal, $\log(N)=x$, which represents a perfect fit.

It turns out that the invariants $\ell_i$ show relatively little variation with increasing Picard number $h$. If we insert typical values for these quantities, together with $h_{\rm CICY}=19$ and $h_{\rm max}=491$ into the above fit result the extrapolated numbers become
\begin{equation}
 N(h_{\rm CICY})\simeq 10^{23}\;,\qquad N(h_{\rm max})\simeq 10^{723}\; ,
\end{equation}
which is not too far away from the earlier result~\eqref{extra1}.

\section{Conclusions}

The fit illustrated in Fig.~\ref{fig4} looks rather convincing and we believe that an extrapolation to large $h=h^{1,1}$ is trustworthy (since the invariants $\ell_i$ show little variation relative to $h$), as long as the underlying model-building assumptions continue to be satisfied for large $h$. 
We believe that this is the case for the CICY dataset and, hence, the number of $\simeq 10^{23}$ standard models within this set should be taken seriously. 

The extrapolation to $h_{\rm max}=491$, the maximal known Picard number of any CY, is more questionable. Some of the model building assumptions made here have not yet been checked for the KS set, and there are even indications that they may not be satisfied. First, it is not clear that the KS set contains manifolds that admit freely-acting symmetries with the same frequency as CICYs. The only systematic checks  carried out for low $h^{1,1}$ where the frequency of symmetries is lower than the CICYs~\cite{Braun:2017juz}.  No information is available for large $h^{1,1}$ yet. 

Another generic feature of CICY models is the frequent absence of large numbers of vector-like pairs, so that checking the index was sufficient to guarantee the correct spectrum for a significant fraction of the models. It is not clear that this feature persists for constructions based on other CY manifolds. In fact, the results of Ref.~\cite{Braun:2017feb} suggest that the presence of phenomenologically problematic numbers of vector-like pairs might be a generic feature of some other CY constructions. Again, no definite statement on this is available for the KS set. 

In summary, the number of $\simeq 10^{723}$ standard models for the extrapolation to $h^{1,1}=491$ should be viewed with considerable caution. 
However, the extrapolation to $h^{1,1}=19$, the maximal Picard number in the CICY set has to be taken seriously and leads to $\simeq 10^{23}$ standard models. Even this number, almost certainly a conservative lower bound, is frighteningly large and beyond current computer storage and systematic search.


\begin{thebibliography}{0}%
\makeatletter
\providecommand \@ifxundefined [1]{%
 \@ifx{#1\undefined}
}%
\providecommand \@ifnum [1]{%
 \ifnum #1\expandafter \@firstoftwo
 \else \expandafter \@secondoftwo
 \fi
}%
\providecommand \@ifx [1]{%
 \ifx #1\expandafter \@firstoftwo
 \else \expandafter \@secondoftwo
 \fi
}%
\providecommand \natexlab [1]{#1}%
\providecommand \enquote  [1]{``#1''}%
\providecommand \bibnamefont  [1]{#1}%
\providecommand \bibfnamefont [1]{#1}%
\providecommand \citenamefont [1]{#1}%
\providecommand \href@noop [0]{\@secondoftwo}%
\providecommand \href [0]{\begingroup \@sanitize@url \@href}%
\providecommand \@href[1]{\@@startlink{#1}\@@href}%
\providecommand \@@href[1]{\endgroup#1\@@endlink}%
\providecommand \@sanitize@url [0]{\catcode `\\12\catcode `\$12\catcode
  `\&12\catcode `\#12\catcode `\^12\catcode `\_12\catcode `\%12\relax}%
\providecommand \@@startlink[1]{}%
\providecommand \@@endlink[0]{}%
\providecommand \url  [0]{\begingroup\@sanitize@url \@url }%
\providecommand \@url [1]{\endgroup\@href {#1}{\urlprefix }}%
\providecommand \urlprefix  [0]{URL }%
\providecommand \Eprint [0]{\href }%
\providecommand \doibase [0]{http://dx.doi.org/}%
\providecommand \selectlanguage [0]{\@gobble}%
\providecommand \bibinfo  [0]{\@secondoftwo}%
\providecommand \bibfield  [0]{\@secondoftwo}%
\providecommand \translation [1]{[#1]}%
\providecommand \BibitemOpen [0]{}%
\providecommand \bibitemStop [0]{}%
\providecommand \bibitemNoStop [0]{.\EOS\space}%
\providecommand \EOS [0]{\spacefactor3000\relax}%
\providecommand \BibitemShut  [1]{\csname bibitem#1\endcsname}%
\let\auto@bib@innerbib\@empty
\end{thebibliography}%


\begin{thebibliography}{99}
\ifx\doiref\asklfhas\newcommand{\doiref}[2]{\href{http://dx.doi.org/#1}{#2}}\fi
\raggedright 
\ifx\arxivref\asklfhas\newcommand{\arxivref}[2]{\href{http://arxiv.org/abs/#1}{arXiv:#1}}\fi
\raggedright


\bibitem{Candelas:1985en} 
  P.~Candelas, G.~Horowitz, A.~Strominger and E.~Witten,
  Nucl.\ Phys.\ B {\bf 258}, 46 (1985).


\bibitem{Greene:1986ar} 
  B.~Greene, K.~Kirklin, P.~Miron, G.~Ross,
  Phys.\ Lett.\ B {\bf 180}, 69 (1986).



\bibitem{Braun:2005ux} 
  V.~Braun, Y.~H.~He, B.~A.~Ovrut and T.~Pantev,
  Phys.\ Lett.\ B {\bf 618}, 252 (2005),
  [hep-th/0501070].

\bibitem{Braun:2005nv} 
  V.~Braun, Y.~H.~He, B.~A.~Ovrut and T.~Pantev,
  JHEP {\bf 0605}, 043 (2006),
  hep-th/0512177.

\bibitem{Bouchard:2005ag} 
  V.~Bouchard and R.~Donagi,
  Phys.\ Lett.\ B {\bf 633}, 783 (2006),
 hep-th/0512149.

\bibitem{Braun:2006ae} 
  V.~Braun, Y.~H.~He and B.~A.~Ovrut,
  JHEP {\bf 0606}, 032 (2006),
  hep-th/0602073.


\bibitem{Anderson:2009mh} 
  L.~B.~Anderson, J.~Gray, Y.~H.~He and A.~Lukas,
  JHEP {\bf 1002}, 054 (2010),
 arXiv:0911.1569 [hep-th].


\bibitem{Anderson:2007nc} 
  L.~B.~Anderson, Y.~H.~He and A.~Lukas,
  JHEP {\bf 0707}, 049 (2007),
 hep-th/0702210.

\bibitem{Anderson:2011ns} 
  L.~B.~Anderson, J.~Gray, A.~Lukas and E.~Palti,
  Phys.\ Rev.\ D {\bf 84}, 106005 (2011),
 arXiv:1106.4804 [hep-th].


\bibitem{Anderson:2012yf} 
  L.~B.~Anderson, J.~Gray, A.~Lukas and E.~Palti,
  JHEP {\bf 1206}, 113 (2012),
 arXiv:1202.1757 [hep-th].\\
-- 
PoS CORFU {\bf 2011}, 096 (2011).

\bibitem{Anderson:2013xka} 
  L.~B.~Anderson, A.~Constantin, J.~Gray, A.~Lukas and E.~Palti,
  JHEP {\bf 1401}, 047 (2014),
 arXiv:1307.4787 [hep-th].

\bibitem{stats} 
  K.~R.~Dienes,
  Phys.\ Rev.\ D {\bf 73}, 106010 (2006)
  [hep-th/0602286].
\\
K.~R.~Dienes, M.~Lennek, D.~Senechal and V.~Wasnik,
  Phys.\ Rev.\ D {\bf 75}, 126005 (2007)
  [arXiv:0704.1320 [hep-th]].
\\
K.~R.~Dienes and M.~Lennek,
  Phys.\ Rev.\ D {\bf 75}, 026008 (2007)
  [hep-th/0610319].
  
\bibitem{Gmeiner:2005vz} 
  F.~Gmeiner, R.~Blumenhagen, G.~Honecker, D.~Lust and T.~Weigand,
  JHEP {\bf 0601}, 004 (2006)
  [hep-th/0510170].
  
  \bibitem{Douglas:2003um} 
  M.~R.~Douglas,
  JHEP {\bf 0305}, 046 (2003)
  [hep-th/0303194].
    
\bibitem{Constantin:2015bea} 
  A.~Constantin, A.~Lukas and C.~Mishra,
  JHEP {\bf 1603}, 173 (2016),
 arXiv:1509.02729 [hep-th].


\bibitem{cicy}
P.~Candelas, A.~M.~Dale, C.~A.~Lutken, R.~Schimmrigk,
  Nucl.\ Phys.\ B {\bf 298}, 493 (1988).\\
P.~Candelas, C.~A.~Lutken, R.~Schimmrigk,
  Manifolds,'' 
  Nucl.\ Phys.\ B {\bf 306}, 113 (1988).
\\  
M.~Gagnon, Q.~Ho-Kim,
  Mod.\ Phys.\ Lett.\ A {\bf 9} (1994) 2235.

\bibitem{huebsch} 
T.~H\"ubsch,
``Calabi-Yau Manifolds",
World Scientific, 1992.



\bibitem{ks} 
  A.~C.~Avram, M.~Kreuzer, M.~Mandelberg and H.~Skarke,
  Nucl.\ Phys.\ B {\bf 505}, 625 (1997),
 hep-th/9703003.
\\
  M.~Kreuzer and H.~Skarke,
  Adv.\ Theor.\ Math.\ Phys.\  {\bf 4}, 1209 (2002),
hep-th/0002240.



\bibitem{palp}
  M.~Kreuzer and H.~Skarke,
  Comput.\ Phys.\ Commun.\  {\bf 157} (2004) 87,
 math/0204356.
\\
  A.~P.~Braun, J.~Knapp, E.~Scheidegger, H.~Skarke and N.~O.~Walliser,
  arXiv:1205.4147.


\bibitem{Altman:2014bfa} 
  R.~Altman, J.~Gray, Y.~H.~He, V.~Jejjala and B.~D.~Nelson,
  JHEP {\bf 1502}, 158 (2015),
 arXiv:1411.1418 [hep-th].


\bibitem{Slansky:1981yr} 
  R.~Slansky,
  Phys.\ Rept.\  {\bf 79}, 1 (1981).

\bibitem{Constantin:2018hvl}
A.~Constantin, and A.~Lukas 
arXiv:1808.09992.

\bibitem{Klaewer:2018sfl}
D.~Klaewer, and L.~Schlechter,
arXiv:1809.02547 [hep-th].

\bibitem{Constantin:2013}
A.~Constantin, 
  arXiv:1808.09993 [hep-th]
  
\bibitem{Buchbinder:2013dna}
E.~I. Buchbinder, A.~Constantin, and A.~Lukas, 
 {JHEP} {\bf
  1403} (2014) 025, arXiv:1311.1941 [hep-th].

\bibitem{Buchbinder:2014qda}
E.~I. Buchbinder, A.~Constantin, and A.~Lukas, 
  {JHEP} {\bf 1406} (2014)
  100, arXiv:1404.2767 [hep-th].

\bibitem{Buchbinder:2014sya}
E.~I. Buchbinder, A.~Constantin, and A.~Lukas, 
{Phys. Lett.} {\bf B748} (2015)
  251--254, arXiv:1409.2412 [hep-th].
  
\bibitem{HeathBrown}
T.~D.~Browning, D.~R.~Heath-Brown, P.~Salberger,
arXiv:math/0410117.


\bibitem{Braun:2017juz}
  A.~Braun, A.~Lukas and C.~Sun,
  Commun.\ Math.\ Phys.\  {\bf 360} (2018) no.3,  935,
 arXiv:1704.07812.
  
\bibitem{Braun:2017feb}
  A.~P.~Braun, C.~R.~Brodie and A.~Lukas,
  JHEP {\bf 1804} (2018) 087,
 arXiv:1706.07688.


\bibitem{He:2013ofa} 
  Y.~H.~He, S.~J.~Lee, A.~Lukas and C.~Sun,
  JHEP {\bf 1406}, 077 (2014),
arXiv:1309.0223 [hep-th].


\bibitem{Braun:2010vc} 
  V.~Braun,
  JHEP {\bf 1104}, 005 (2011),
 arXiv:1003.3235 [hep-th].

\bibitem{Lukas:2017vqp} 
  A.~Lukas and C.~Mishra,
arXiv:1708.08943.

\bibitem{Candelas:2017ive} 
  P.~Candelas and C.~Mishra,
  Fortsch.\ Phys.\  {\bf 66}, 1800017 (2018),
arXiv:1709.01081 [hep-th].


\bibitem{Candelas:2016fdy} 
  P.~Candelas, A.~Constantin and C.~Mishra,
  Fortsch.\ Phys.\  {\bf 66}, 1800029 (2018),
 arXiv:1602.06303 [hep-th].


\bibitem{Candelas:2008wb} 
  P.~Candelas and R.~Davies,
  Fortsch.\ Phys.\  {\bf 58}, 383 (2010),
 arXiv:0809.4681 [hep-th].\\
 
 P.~Candelas and A.~Constantin, 
 {Fortsch.Phys.} {\bf 60}
  (2012) 345--369, arXiv:010.1878.\\
 
  P.~Candelas, A.~Constantin and C.~Mishra,
  Fortsch.\ Phys.\  {\bf 64}, no. 6-7, 463 (2016),
 arXiv:1511.01103.

\bibitem{Candelas:2007ac} 
  P.~Candelas, X.~de la Ossa, Y.~H.~He and B.~Szendroi,
  Adv.\ Theor.\ Math.\ Phys.\  {\bf 12}, no. 2, 429 (2008),
 arXiv:0706.3134.
 
 
\bibitem{Constantin:2016xlj}
A.~Constantin, J.~Gray, and A.~Lukas, 
 {\em JHEP} {\bf 01} (2017) 001, arXiv:1607.01830.

 
    

\end{thebibliography}
\end{document}